\documentclass[a4paper,11pt]{article}
\usepackage{jcappub2} 
\usepackage{lineno}

\def\ba{\begin{eqnarray}}
\def\ea{\end{eqnarray}}

\usepackage{color}

\arxivnumber{2601.18423 [hep-ph]}
\title{\boldmath Electric and magnetic timelike form factors of hyperons
at large transfer momentum}

\author[a]{G.~Ramalho,}
\author[b,c]{M.~T.~Pe\~na,}
\author[d]{K.~Tsushima,}
\author[a]{and Myung-Ki Cheoun}
\affiliation[a]{Department of Physics and OMEG Institute, Soongsil University, \\
  Seoul 06978, Republic of Korea}
\affiliation[b]{LIP, Laborat\'orio de Instrumenta\c{c}\~ao e F\'{i}sica 
Experimental de Part\'{i}culas, \\
Avenida Professor Gama Pinto, 1649-003 Lisboa, Portugal}
\affiliation[c]{Departamento de F\'{i}sica e Departamento de Engenharia e Ci\^encias Nucleares, \\
Instituto Superior T\'ecnico, 
Universidade de Lisboa, \\
Avenida Rovisco Pais,  1049-001 Lisboa, Portugal}
\affiliation[d]{Laborat\'orio de 
  F\'{i}sica Te\'orica e Computacional -- LFTC, \\
  Programa de P\'osgradua\c{c}\~ao em F\'{i}sica Computacional, 
Universidade Cidade de  S\~ao Paulo,  \\
01506-000,  S\~ao Paulo, SP, Brazil}

\emailAdd{gilberto.ramalho2013@gmail.com}

\abstract{There has been considerable progress in the study of the electromagnetic form factors of baryons in the timelike region, through electron-positron scattering reactions ($e^+ e^- \to B \bar B$), in the last two decades.
 Timelike experiments reveal information about the distribution of charge and magnetism inside the hyperons that cannot be obtained in spacelike experiments (electron scattering on baryons). Motivated by the novel data, we extend to the timelike region, without any further parameter fitting, a covariant quark model developed for the spacelike region that takes into account the meson cloud excitations of the baryon cores. We use the formalism to calculate the electric ($G_E$) and magnetic ($G_M$) form factors of spin 1/2 baryons in the large square transfer momentum $q^2$ region. Our calculations are compared with the available data from CLEO and BESIII above $q^2=10$ GeV$^2$. We conclude that our predictions for the effective form factors (combination between $G_E$ and $G_M$) are in good agreement with the $q^2 > 15$ GeV$^2$ data for $\Lambda$, $\Sigma^+$, $\Sigma^0$, $\Xi^-$ and $\Xi^0$.
 Upcoming data for $\Sigma^-$ can be used to further test our predictions.
 We also compare our model calculations with the available data for ratio $|G_E/G_M|$.
 We conclude that the present $q^2$ data range is not large enough to test our calculations, but that a more definitive test can be performed by upcoming data above $q^2=20$ GeV$^2$.}

\begin{document}
\maketitle
\flushbottom

\section{Introduction}

The study of the structure of the baryons has been dominated by spacelike experiments based on the scattering of electrons on nucleon targets (square transfer momentum $q^2=$ $-Q^2 \le 0$) in facilities like Jefferson Lab, MAMI, and MIT-Bates~\cite{NSTAR}.
In recent years, there was significant progress in the experimental study of the electromagnetic structure of the baryons in the timelike region (invariant square transfer momentum $q^2 > 0$), based on electron-positron collisions in facilities like BaBar, CLEO, Belle and BES~\cite{Hyperons12,Review11,Review12,Dobbs17a,BES-CompletM,Review2}.
The information about the electromagnetic structure of spin 1/2 baryons $B$ can be obtained from the differential and total cross sections on the annihilation reactions $e^+ e^- \to B \bar B$ in a region where the square transfer momentum is $q^2 =s \ge 4 M_B^2$ ($M_B$ is the mass of the baryon).
In the timelike region, the electromagnetic form factors are complex functions of $q^2$.
From the experiments we obtain the modulus of the electric ($G_E$) and magnetic ($G_M$) form factors, that can be combined to define the effective form factor $|G (q^2)|$, and the electromagnetic ratio $R(q^2)$ [$\tau =q^2/(4M_B^2)$]~\cite{Hyperons12,Review11,Review12}
\ba
|G(q^2)|^2 = \frac{2 \tau |G_M(q^2)|^2 + |G_E(q^2)|^2}{1+ 2 \tau},
\hspace{1.5cm}
R(q^2) = \frac{|G_E(q^2)|}{|G_M(q^2)|}.
\ea
These experiments reveal information about the distribution of charge and magnetism inside the hyperons that cannot be obtained by spacelike experiments (short lifetime of hyperons).
The effective form factors $|G|$ have been measured for several baryons at BaBar, CLEO, BESIII and Belle.
Of interest are the measurements at CLEO and BESIII above $q^2=10$ GeV$^2$~\cite{Dobbs17a,BESIII-Data1,BESIII-Data2}.

In the last few years, it was also possible to measure the polarization of the baryon and antibaryon final states, allowing the determination of the relative phase between $G_E$ and $G_M$: $\Delta \Phi$ at BESIII for the $\Lambda$ and $\Sigma^+$~\cite{BES-CompletM}.
With the knowledge of $|G_M|$, $|G_E|$ and $\Delta \Phi$, (also function of $q^2$), one can write:
\ba
\frac{G_E}{G_M} = \frac{|G_E|}{|G_M|} e^{i \Delta \Phi},
\hspace{2cm}
{\rm Re}  \left( \frac{G_E}{G_M} \right)
    = 
    \frac{|G_E|}{|G_M|} \cos \Delta \Phi.
    \label{eqGEGM}
    \ea
    
Motivated by the experimental measurements on $|G|$ and $R$, we use a covariant quark model~\cite{Spectator}, developed for the study of electromagnetic transitions between baryon states in the spacelike region, to make predictions for the large $q^2$ timelike region~\cite{Hyperons12}.
The model takes into account both the effects of valence quarks and the excitations of the meson cloud, which dresses the bare baryons.

The extension to the timelike region is made considering asymptotic relations valid in the large $q^2$ region, based on unitarity and analyticity.
For arbitrarily large values of $|q^2|= |Q^2|$ we can use $G_\ell (q^2) = G_\ell^{\rm SL} (q^2)$, for $\ell =E,M$, to determine the form factors $G_\ell (q^2)$ in the timelike region, based on the spacelike form factors $G_\ell^{\rm SL} (Q^2)$, defined in terms of $Q^2= -q^2$ for the region $Q^2 \ge 0$~\cite{Review11}.

 However, one notices that the asymptotic relations are strictly valid for the limit $q^2 \to \infty$, and it is not clear that $q^2=0$ should be the center of the reflection relation between spacelike and timelike form factors, since the threshold of the timelike form factors is $q^2= 4 M_B^2$.

In order to take into account this ambiguity, we consider a finite correction to the variable $q^2$, using $q^2 \to q^2 - 2 M_B^2$, shifting the center of the reflection to the point between the two thresholds ($\ell =E,M$)~\cite{Hyperons12}
\ba
G_\ell (q^2) \equiv G_\ell^{\rm SL} (q^2 - 2 M_B^2).
\ea
The theoretical uncertainties associated with the finite corrections to the asymptotic relations can be estimated using the limits determined by the calculation centered on the spacelike threshold, $G_\ell (q^2) = G_\ell^{\rm SL} (q^2)$, and the timelike threshold, $G_\ell (q^2) = G_\ell^{\rm SL} (q^2- 4 M_B^2)$.
When $q^2$ is very large, the theoretical deviations become very small and the upper and lower estimates of the form factors converge to the same value, providing accurate predictions that can be tested by experimental data.
The limits of deviation from the asymptotic relations are included in the following figures.

\section{Numerical results}

We use the covariant spectator quark model~\cite{Spectator}.
The formalism was developed originally for the study of transitions of the nucleon and nucleon resonances~\cite{nstars} and transitions between baryon states in the spacelike region ($Q^2 = - q^2 \ge 0$)~\cite{Octet4,OctetDecuplet}, but has also been used in the study of inelastic timelike transitions~\cite{Dalitz}, and the electroweak structure of baryons in vacuum and in a nuclear medium~\cite{Axial-Medium} and the nucleon deep inelastic scattering~\cite{NucleonDIS}.

Within the formalism the electromagnetic current
from transitions between baryon states are determined
using relativistic impulse approximation applied 
to the photon-quark vertex, and the baryon wave functions are reduced
to quark-diquark components, where the diquark is on shell~\cite{Spectator,Octet4,OctetDecuplet}.
The quark internal structure associated with the gluon and quark-antiquark dressing is parametrized by constituent quark form factors.
The quark-diquark radial wave functions are determined by fitting lattice QCD data, based on the SU(3) flavor symmetry structure, that is broken explicitly~\cite{Octet4}.
In the low-$Q^2$ region, we consider also effective parametrizations of the meson cloud dressing of the baryon states~\cite{Dalitz,OctetDecuplet}.
In the calculations, we use the parameters determined by the model from Refs.~\cite{Octet4,OctetDecuplet}.

\begin{figure*}[t]
  \vspace{.1cm}
\centerline{
\mbox{
\includegraphics[width=1.9in]{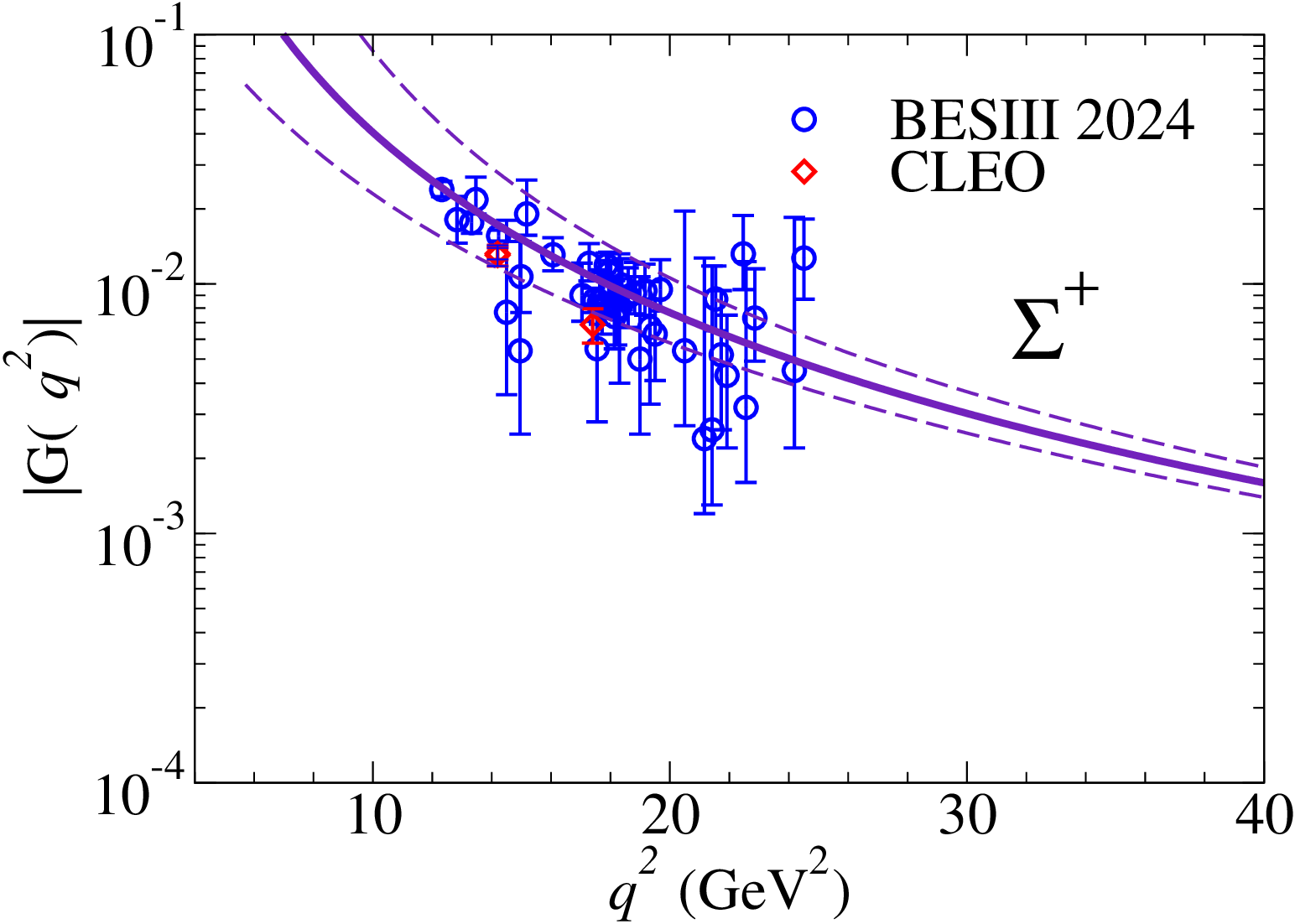} \hspace{.1cm}
\includegraphics[width=1.9in]{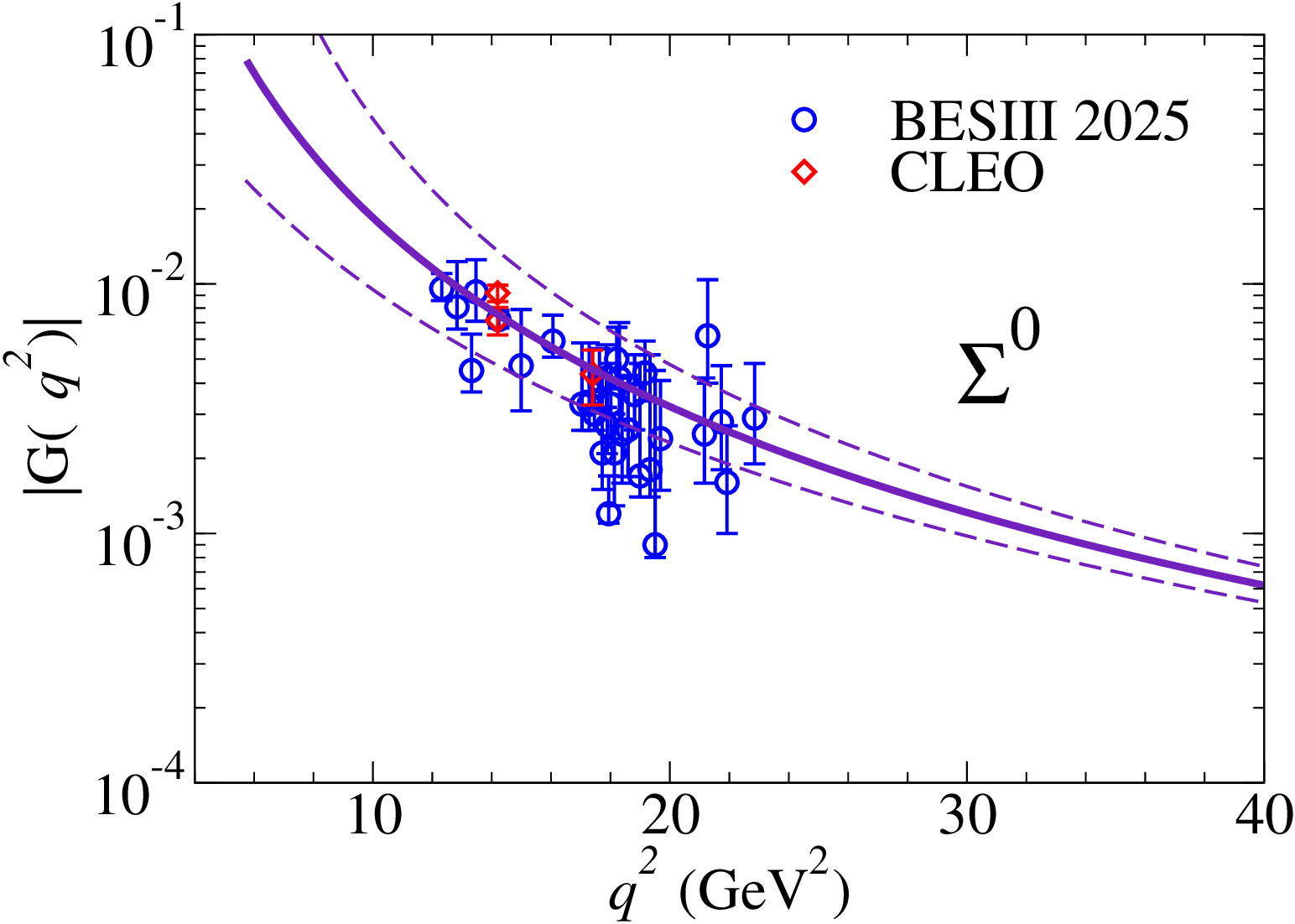} \hspace{.1cm}
\includegraphics[width=1.9in]{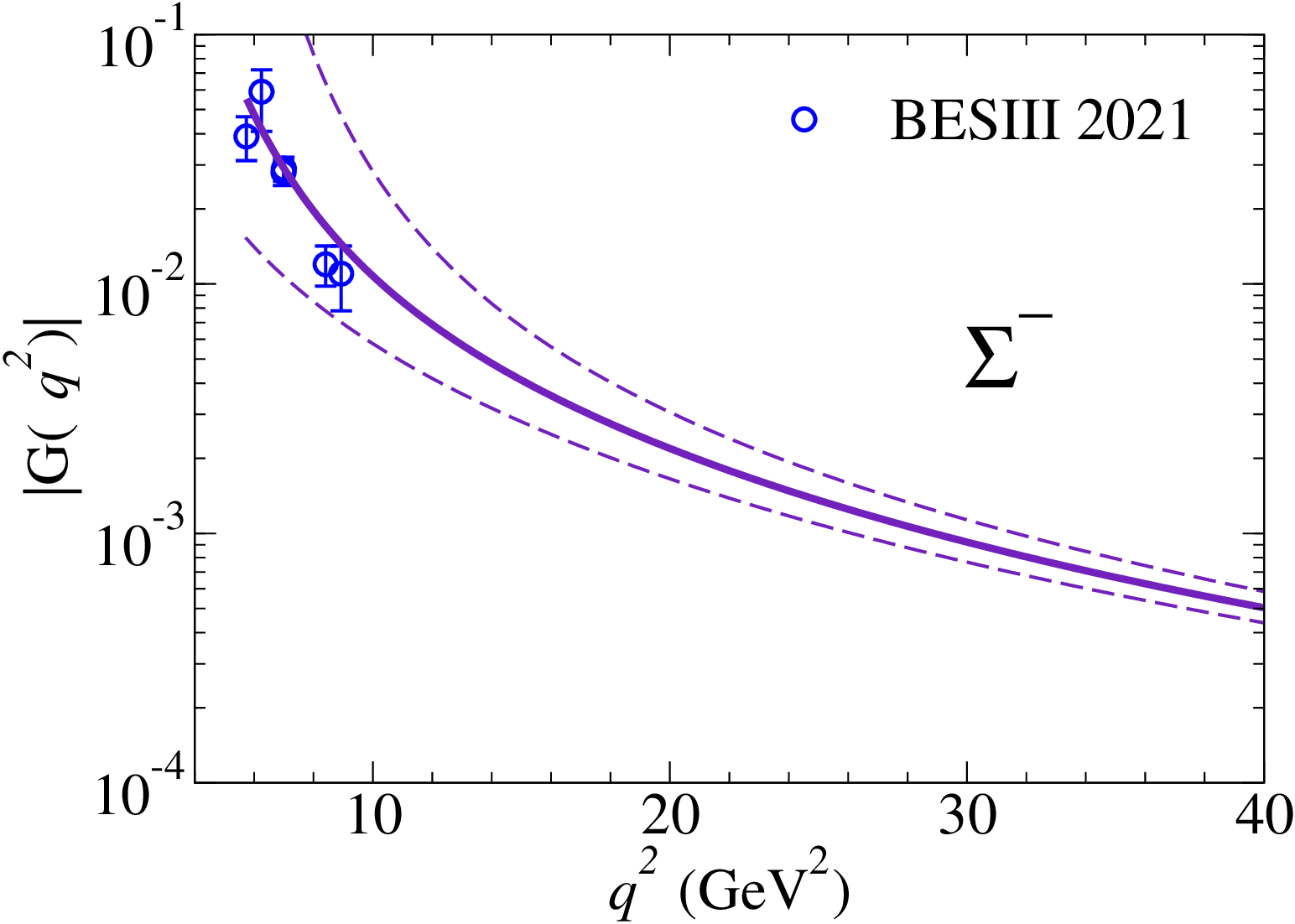}
}}
\vspace{.15cm}
\centerline{
\mbox{
\includegraphics[width=1.9in]{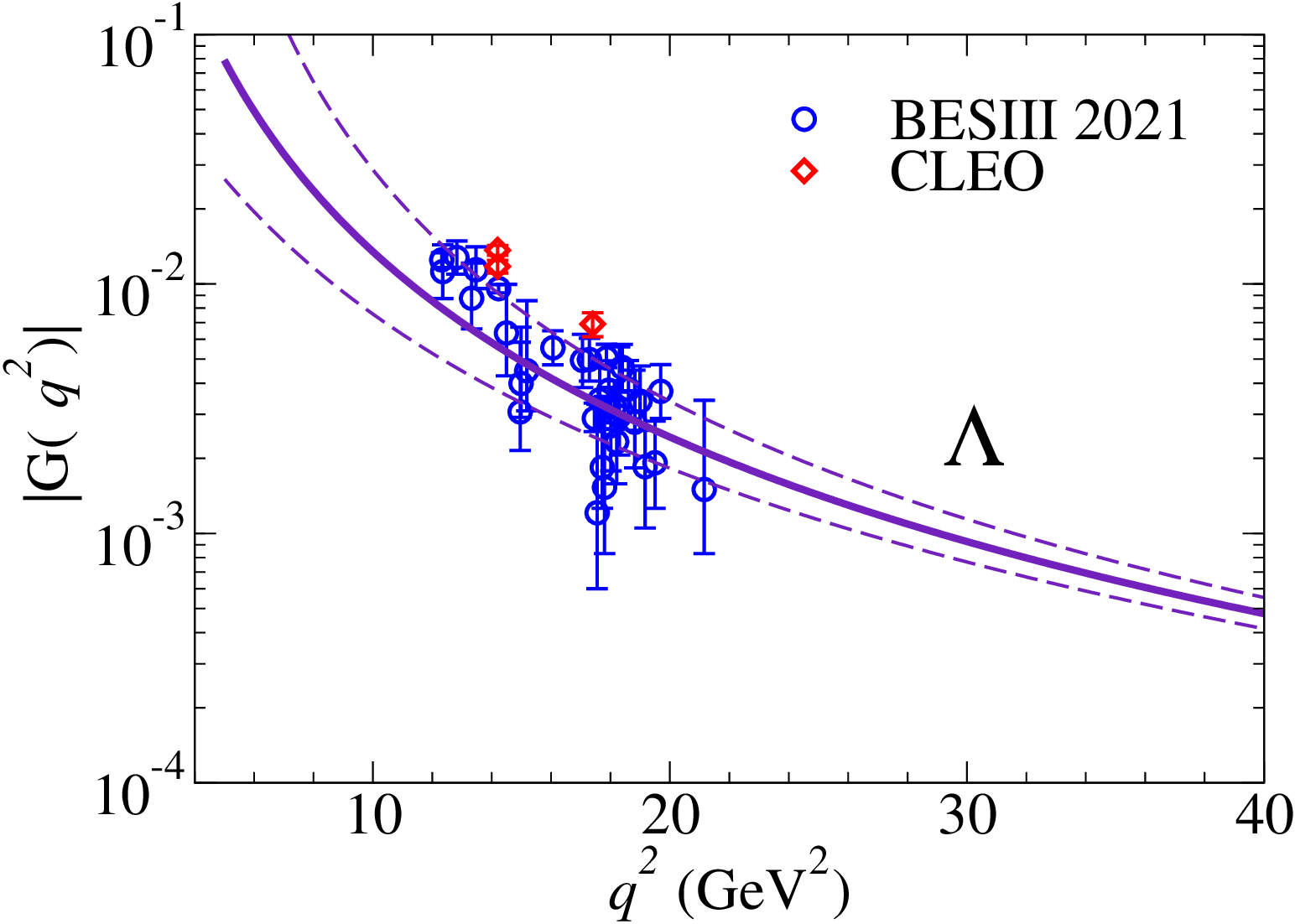} \hspace{.1cm}
\includegraphics[width=1.9in]{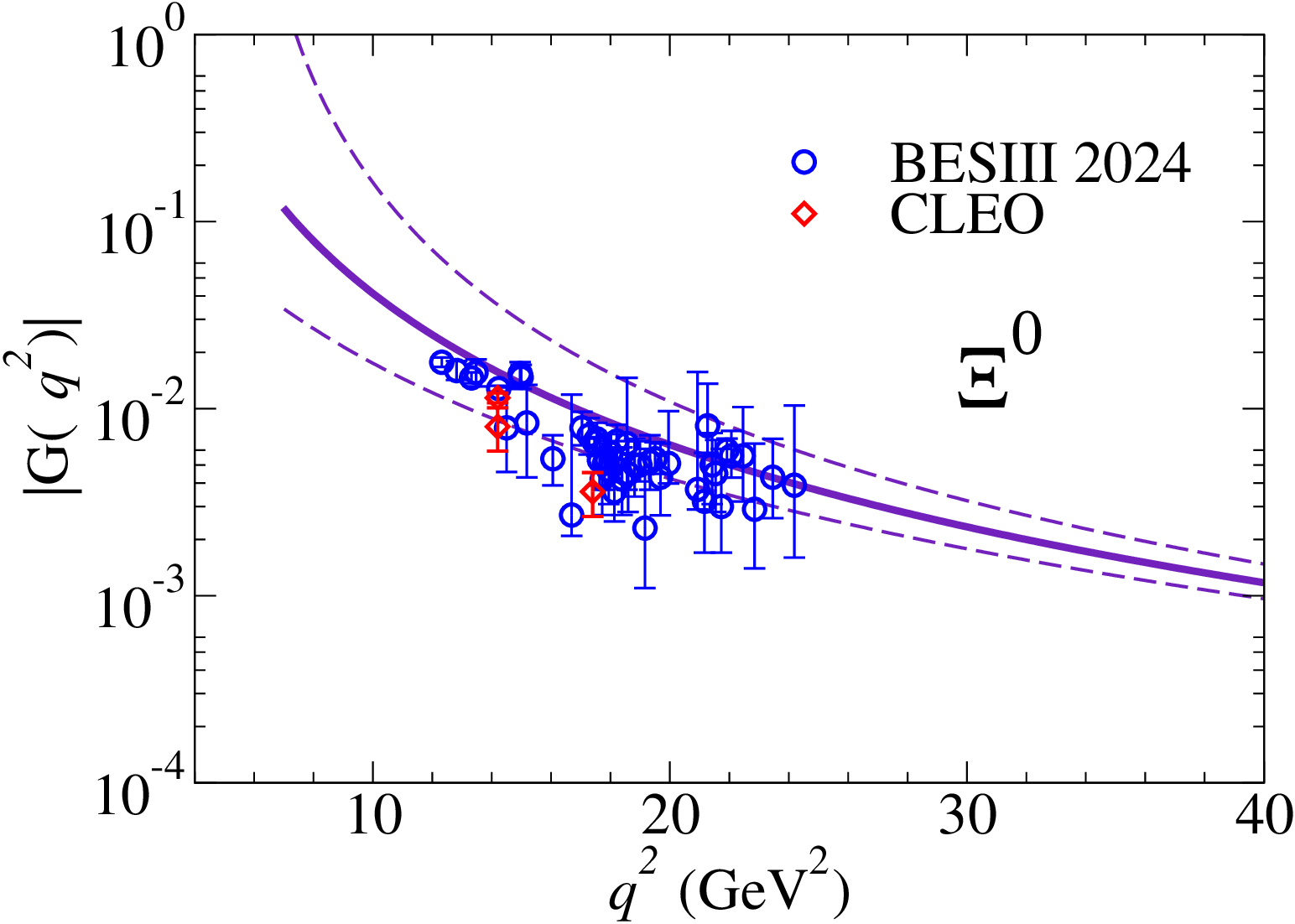} \hspace{.1cm}
\includegraphics[width=1.9in]{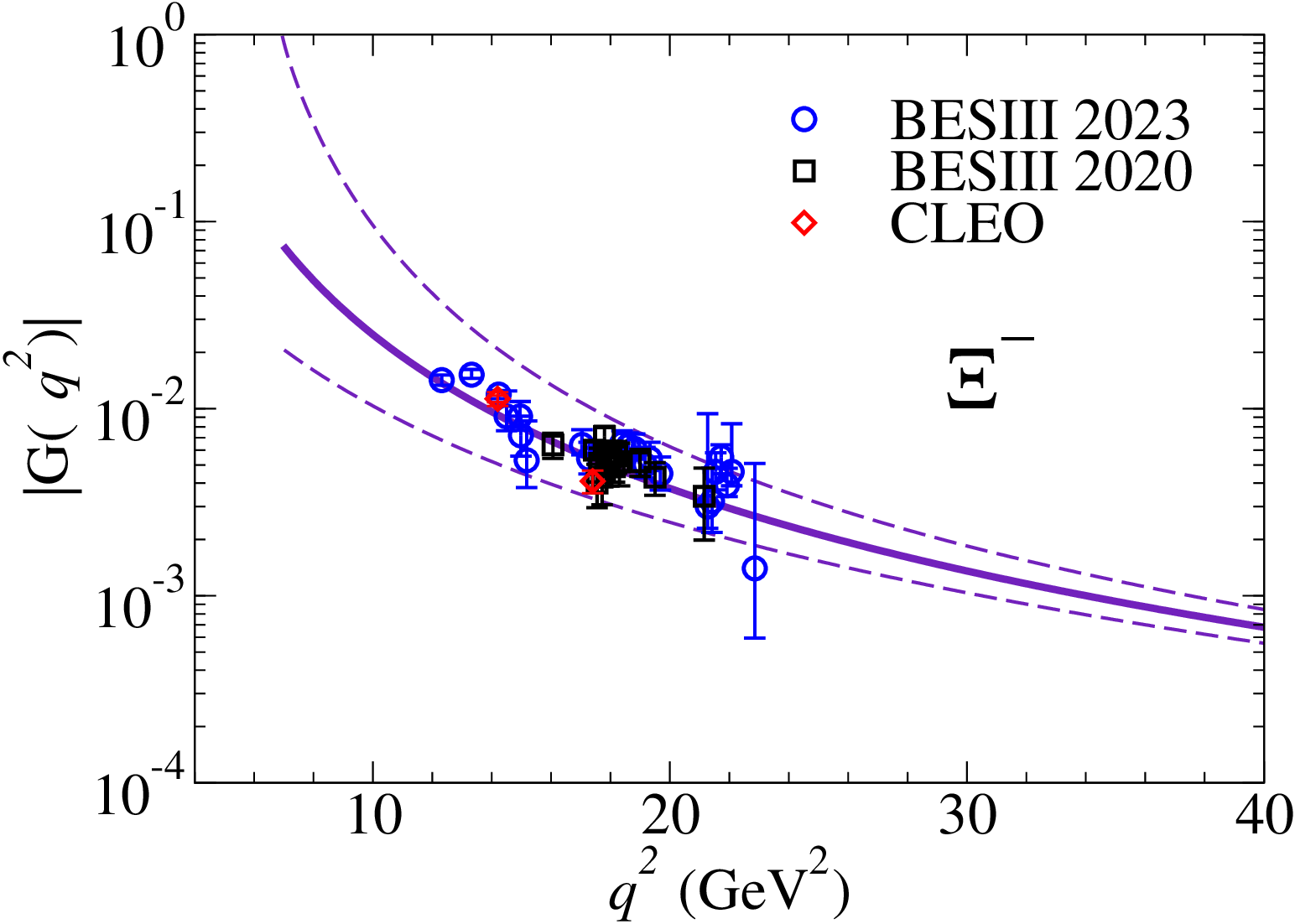}
}}
\caption{\footnotesize{Effective form factors $|G(q^2)|$ of the hyperons for $q^2 > 10$ GeV$^2$ (solid line)~\cite{Hyperons12}.
    The dashed lines represent the limits of deviations from the asymptotic relations.
    The data are from CLEO~\cite{Dobbs17a} and BESIII~\cite{BESIII-Data1,BESIII-Data2}.
    For $\Sigma^-$ we include also BESIII data below $q^2=10$ GeV$^2$.
} \label{figG1}}
\end{figure*}

In Figure~\ref{figG1}, we compare our model calculations~\cite{Hyperons12} for $|G|$ with the large $q^2$ data for the hyperons, including the limits of the theoretical estimates.
Our predictions, performed before the measurements from BESIII, are in very good agreement with the BESIII and CLEO data above $q^2=15$ GeV$^2$.
Our predictions for $\Sigma^-$ are  waiting for measurements at BES for  $q^2 > 10$ GeV$^2$.
Our calculations for $|G|$ suggest that the region $q^2 > 15$ GeV$^2$ is already in the range where the asymptotic behavior of the form factors can be observed.
We notice, however, that we may still be in the nonperturbative QCD region, and that the onset for the perturbative QCD falloff of the form factors happens in a much higher region of $q^2$~\cite{Hyperons12}.
We do not discuss the oscillatory behavior observed on the form factors of baryons because the effects are expected to be suppressed at large $q^2$~\cite{Oscilations}.

\begin{figure*}[t]
  \vspace{.3cm}
  \centerline{\mbox{
\includegraphics[width=2.9in]{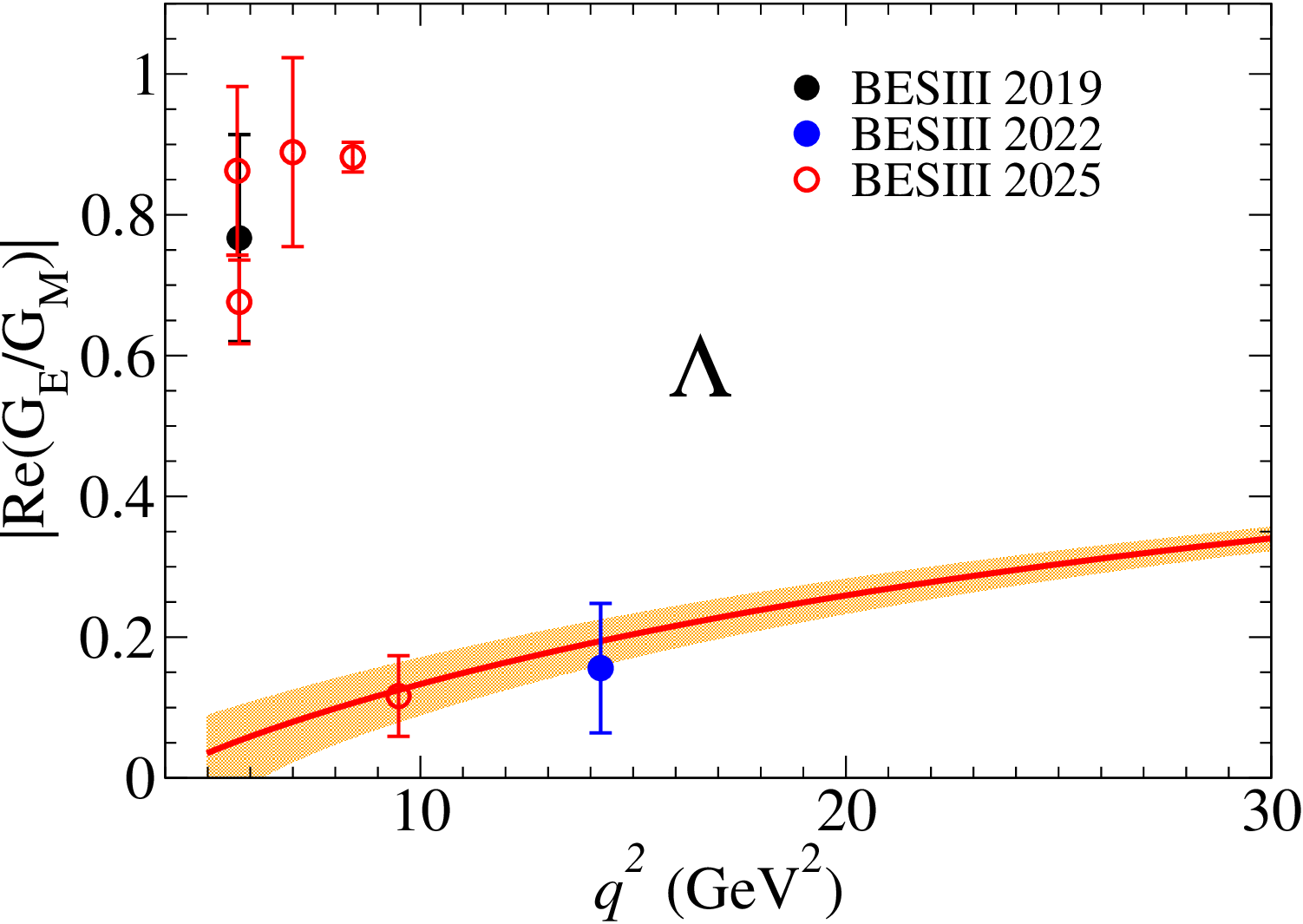} \hspace{.1cm}
\includegraphics[width=2.9in]{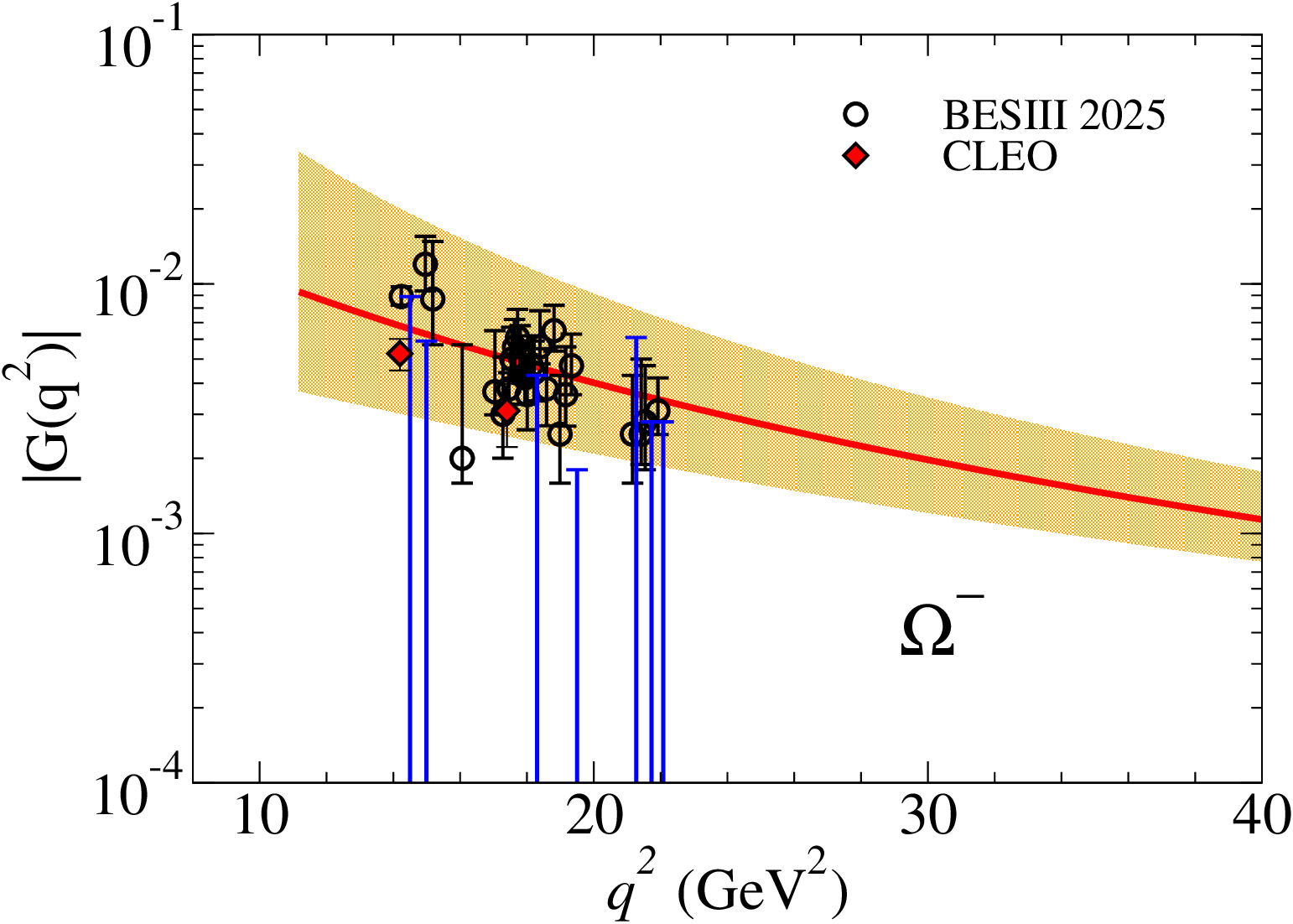} }}
  \caption{\footnotesize{ {\bf At the left:}
        Real part of the ratio $G_E/G_M$ for the $\Lambda$ with theoretical uncertainties. The data are from Refs.~\cite{BESIII22-25}.
        {\bf At the right:}
        Effective form factor of the $\Omega^-$~\cite{Omega3}.
        The data are from  BESIII~\cite{BESIII-Omega} and CLEO~\cite{Dobbs17a}.
        The blue lines indicate the upper limit of the measurements $|G|=0$.
        \label{figRatio1}}}
\end{figure*}

We also make predictions for the ratio $|G_E/G_M|$~\cite{Hyperons12}, that has been measured recently for $\Lambda$ and $\Sigma^+$ for moderated values of $q^2$ ($q^2 \sim 10$ GeV$^2$)~\cite{BES-CompletM,BESIII22-25}.
Our model calculations are expected to be accurate only for larger values of $q^2$, since they are based on asymptotic relations valid for a region where the imaginary parts of the form factors can be neglected.
Our model calculations underestimate the available data~\cite{Hyperons12}, but we can also compare our calculations directly with the real part of $G_E/G_M$, when the relative phases are known [see Eq.~(\ref{eqGEGM})].
Our estimates for $\Lambda$ are compared with the data on the left side of Figure~\ref{figRatio1}, showing that our model is consistent with the data above $q^2=8$ GeV$^2$.
These results suggest that $G_E/G_M$ became real near $q^2 \simeq 10$ GeV$^2$, consistently with the calculations from Refs.~\cite{BESIII22-25,BESIII-Lambda-2025b}.
Overall, our calculations hint that the onset of validity of our model for $|{\rm Re}(G_E/G_M)|$ is not so far from the region $q^2=20$--30 GeV$^2$, to be reached soon~\cite{Hyperons12}.


The calculations of the effective form factors can be extended to baryons with spin 3/2~\cite{Hyperons12,Omega3}.
Our model for the $\Omega^-$ is the result of a global fit to the $\Omega^-$ electromagnetic form factor data dominated by the spacelike region (96\%)~\cite{Omega3}.
The results are presented on the right side of Figure~\ref{figRatio1} in comparison with the recent BESIII data~\cite{BESIII-Omega}.
Although the range of the available data is not very high ($q^2 \simeq 5 M_\Omega^2$--$8 M_\Omega^2$), we can notice a good agreement with the data within the model uncertainties.

\section{Outlook and conclusions}

We present predictions for the elastic form factors of the hyperons for the large $q^2$ timelike region, based on the covariant spectator quark model, without any parameter fitting.
Our calculations are based on a model developed successfully for the study of baryons in the spacelike region.
The formalism is extended to the timelike region using asymptotic relations valid for the large-$q^2$ region, and the theoretical uncertainties are also quantified.

Our estimates for the effective form factor $|G|$ are in good agreement with recent BESIII data for $\Lambda$, $\Sigma^+$, $\Sigma^0$, $\Xi^0$, and $\Xi^-$ above $q^2= 15$ GeV$^2$.
To be tested in the future are predictions for $\Sigma^-$ above $q^2= 10$ GeV$^2$. 
As for the ratio $|G_E/G_M|$, measurements for higher values of $q^2$ are necessary to test our model calculations.
There are, however, signs that our predictions may be in good agreement with the data for $|{\rm Re}(G_E/G_M)|$ for not so large values of $q^2$, providing that the measurements of the  $\Delta \Phi$ are accurate.
Precise data from BESIII, Belle or PANDA for $|G|$, $|G_E/G_M|$ and  $\Delta \Phi$ may be used in the near future to test our predictions.

Our formalism may be extended in the future to the description of inelastic timelike transitions, such as, $e^+ e^- \to \Lambda \bar \Sigma^0$, $\Sigma^0 \bar \Lambda$ and baryons with heavy quarks.

\acknowledgments

G.R.~and M.-K.C.~were supported by the National
Research Foundation of Korea (Grant  No.~RS-2021-NR060129).
M.T.P.~was supported by the Portuguese Science Foundation FCT
under project CERN/FIS-PAR/0023/2021, and FCT computing project 2021.09667.
K.T.~was supported by CNPq, Brazil, Processes 
No.~304199/2022-2, FAPESP Process No.~\\2023/07313-6,
and by Instituto Nacional de Ci\^{e}ncia e Tecnologia -- Nuclear Physics and Applications
(INCT-FNA), Brazil, Process No.~464898/2014-5.





\end{document}